\documentclass[aps,prl,reprint,superscriptaddress,floatfix,amsmath,showpacs]{revtex4-1}

\usepackage{graphicx}
\usepackage{amsmath}
\usepackage{textcomp}
\usepackage{xcolor}

\usepackage{hyphenat}
\usepackage[colorlinks=true,allcolors=blue]{hyperref}

\usepackage{xcolor}
\usepackage[normalem]{ulem}
\usepackage{graphicx}

\definecolor{g-blue}{rgb}{0.83,0.95,1}
\definecolor{g-yellow}{rgb}{1,1,0.7}
\definecolor{g-green}{rgb}{0.9,1,0.9}
\definecolor{green}{rgb}{0,0.6,0}
\definecolor{cyan}{rgb}{0,0.7,0.7}
\definecolor{black}{rgb}{0,0,0}
\definecolor{grey}{rgb}{0.4 ,0.4 ,0.4 }

\usepackage{bm}
\usepackage{amssymb}
\usepackage{amsfonts}
\usepackage{amsmath}
\usepackage{graphicx}
\usepackage{amsmath,bm,epsfig}

\renewcommand{\sb}[1]{_{\text {#1}}}  
\def\Sb#1{_{\scriptscriptstyle\rm{#1}}}

\begin{document}
	
	\title{Long-distance supercurrent transport in a room-temperature \\ Bose-Einstein magnon condensate}
	
	\author{Dmytro~A.~Bozhko}
	\affiliation{Fachbereich Physik and Landesforschungszentrum OPTIMAS, Technische Universit\"at Kaiserslautern, 67663 Kaiserslautern, Germany \looseness=-1}
	
	\author{Alexander~J.~E.~Kreil}
	\affiliation{Fachbereich Physik and Landesforschungszentrum OPTIMAS, Technische Universit\"at Kaiserslautern, 67663 Kaiserslautern, Germany \looseness=-1}
	
	\author{Halyna\,Yu.\,Musiienko-Shmarova}
	\affiliation{Fachbereich Physik and Landesforschungszentrum OPTIMAS, Technische Universit\"at Kaiserslautern, 67663 Kaiserslautern, Germany \looseness=-1}
	
	\author{Alexander~A.~Serga}
	\affiliation{Fachbereich Physik and Landesforschungszentrum OPTIMAS, Technische Universit\"at Kaiserslautern, 67663 Kaiserslautern, Germany \looseness=-1}
	
	\author{Anna~Pomyalov}
	\affiliation{Department of Chemical and Biological Physics, Weizmann Institute of Science, Rehovot 76100, Israel \looseness=-1}
	
	\author{Victor~S.~L'vov}
	\affiliation{Department of Chemical and Biological Physics, Weizmann Institute of Science, Rehovot 76100, Israel \looseness=-1}
	
	\author{Burkard~Hillebrands}
	\email{hilleb@physik.uni-kl.de}
	\affiliation{Fachbereich Physik and Landesforschungszentrum OPTIMAS, Technische Universit\"at Kaiserslautern, 67663 Kaiserslautern, Germany \looseness=-1}
	
\begin{abstract}
	\noindent\textbf{\nohyphens
	The term supercurrent relates to a macroscopic dissipation-free collective motion of a quantum condensate and is commonly associated with such famous low-temperature phenomena as superconductivity and superfluidity. Another type of motion of quantum condensates is second sound---a wave of the density of a condensate. Recently, we reported on an enhanced decay of a parametrically induced Bose-Einstein condensate (BEC) of magnons caused by a supercurrent outflow of the BEC phase from the locally heated area of a room temperature magnetic film. Here, we present the direct experimental observation of a long-distance spin transport in such a system. The condensed magnons being pushed out from the potential well within the heated area form a density wave, which propagates through the BEC many hundreds of micrometers in the form of a specific second sound pulse---Bogoliubov waves---and is reflected from the sample edge. The discovery of the long distance supercurrent transport in the magnon BEC further advances the frontier of the physics of quasiparticles and allows for the application of related transport phenomena for low-loss data transfer in perspective magnon spintronics devices. \looseness=-1   
		}
\end{abstract}	
	\maketitle
	
	\everypar{\looseness=-1}
	

Supercurrent is a macroscopic quantum phenomenon when many bosons (real- or quasiparticles) being self-assembled in one quantum state with minimum energy and zero velocity---a Bose-Einstein condensate (BEC)\cite{Einstein1924,Einstein1925,Borovik-Romanov1984,Anderson1995,Davis1995,Butov2001,Kasprzak2006,Safonov1989,Demokritov2006,Balili2007,Klaers2010}---move as a whole due to a phase gradient imposed on their joint wave function.  This phenomenon being mostly associated with resistant-free electric currents of Cooper pairs\cite{Bardeen1957} in superconductors and superfluidity of liquid Helium \cite{Kapitza1938,Allen1938,Landau1941,Osheroff1972,Leggett2004} is, however, much more widespread \cite{Sonin2010,Page2017,Bunkov2018}. It is experimentally confirmed in the quantum condensates of diluted ultracold gases \cite{Matthews1999, Raman1999}, of nuclear magnons in liquid $^3$He \cite{Borovik-Romanov1988,Bunkov2009,Bunkov2012}, of polaritons in semiconductor microcavities \cite{Amo2009} and, recently, of electron magnons in room-temperature ferrimagnetic films \cite{Bozhko2016}. Supercurrents being topologically confined often manifest themselves in a form of quantum vortices\cite{Matthews1999,Bunkov2008,Nowik-Boltyk2012}.

The quantum condensate supports another form of motion---second sound \cite{Landau1941,Pitaevskii1968}. Second sound can be considered as elementary excitations of various types, which can propagate in continuous media with an almost linear dispersion law in the long-wavelength limit. The term second sound stems from an analogy with the ordinary sound waves or first sound---the wave oscillations of media density and mechanical momentum. The most well-known example of second sound is anti-phase oscillations of the densities $\rho_\mathrm{n}$ and  $\rho_\mathrm{s}$ of the normal-fluid and superfluid components of superfluid $^4$He, in which the total density $\rho = \rho_\mathrm{n} + \rho_\mathrm{s}$ does not oscillate \cite{Landau1941}. These oscillations can be associated with temperature waves, because the ratio $\rho_\mathrm{n} / \rho_\mathrm{s}$ strongly depends on the local temperature, while $\rho$ in $^4$He is practically temperature independent.

Some solid dielectrics represent another system type which supports the propagation of temperature waves at low temperatures \cite{Ackerman1966,McNelly1970,Gurevich1988,Lee2015}. In this case, the second sound is the wave of density of the phonon occupation numbers, with phonon being a quantum of the first sound. 

By exploring the spatio-temporal dynamics of a magnon BEC prepared by microwave parametric pumping in a room-temperature single-crystal film of a magneto-dielectric material \cite{Gurevich1996}, we discovered a new type of second sound, which is the wave excitation of the magnon BEC density and is different from a temperature wave. Furthermore, we demonstrate a strong connection between the supercurrent-type and the second-sound-type  motions of the magnon BEC. In our experiment, a magnon supercurrent flowing out from a thermally induced magnetic inhomogeneity \cite{Bozhko2016} creates perturbations in the BEC density, which travel through the homogeneous BEC areas hundreds of micrometers almost without changing its form. Detailed analysis of the propagation features of these perturbations in the framework of the Gross-Pitaevskii equation allows us to describe the observed phenomenon as a solitary wave of Bogoliubov excitations with oscillations of both the amplitude and the phase of the magnon BEC's wave function. In the long-wavelength limit the Bogoliubov waves have a linear dispersion law and thus, can be considered as a second sound wave featuring long-distance propagation through the magnon condensate.


\begin{figure}[b!]
	\includegraphics[width=1.\columnwidth]{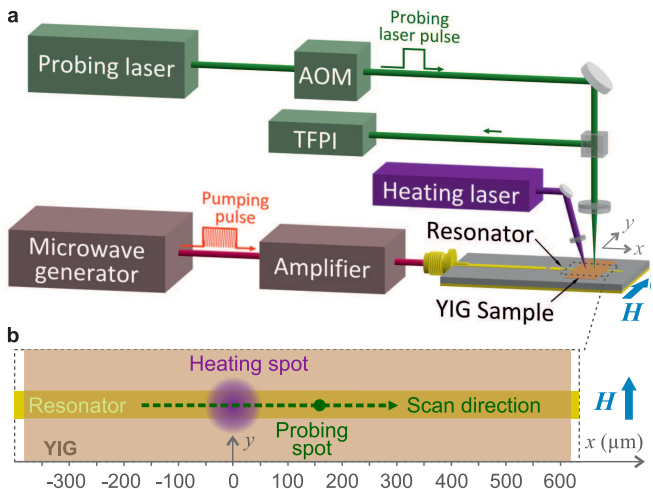}
	\caption{\label{Fig1_Setup} \textbf{Experimental setup.} \textbf{a}, Schematic illustration of the experimental set-up. A probing green laser beam of  $532\,\mathrm{nm}$ wavelength is focused onto a YIG-film sample, which is fixed on top of a pumping microstrip resonator driven by powerful microwave pulses. The probing beam is chopped by an acousto-optic modulator (AOM) to reduce a parasitic heating of the sample. The light and microwave pulses are synchronized allowing for optical observation of the after-pumping evolution of a magnon BEC. The light scattered by magnons is directed to a tandem Fabry-P\'erot interferometer (TFPI) for intensity-, frequency-, and time-domain analysis. A blue laser of $405\,\mathrm{nm}$ wavelength is used for local heating of the sample and is focused into a $80\,\mathrm{\mu m}$ spot in the middle point of the resonator. The blue laser source and the YIG sample are mounted on a single movable stage to hold unchanged a predefined position of the heated area on YIG-film surface in the process of sample motion. The motion of the stage relative to the focal green spot allows for the probing of the magnon gas density in different points of the sample. \textbf{b}, The inset schematically shows a magnified view of the scan area. In order to ensure constant pumping conditions and, thus, spatially uniform density distribution of the magnon BEC in the probing direction, the scan is performed in $x$-direction along the microstrip resonator (across the bias magnetic field $\textbf{\textit{H}}$). 
	}
\end{figure}

\bigskip
{
	\fontsize{11pt}{10.0pt}
	\selectfont
	\noindent\textcolor{black}{\textbf{Experiment}}}

\noindent 
The experiment is carried out in a sample (see Methods) cut out from a single-crystal ferrimagnetic film of Yttrium Iron Garnet (YIG) \cite{Cherepanov1993}. This magneto-dielectric material possesses the lowest known magnetic damping in nature and is one of the favourite magnetic media for fundamental and applied studies in modern magnonics and spintronics \cite{Serga2010}. The detection of magnon dynamics is performed by frequency- and time-resolved Brillouin Light Scattering (BLS) spectroscopy with wavevector sensitivity (see Methods). Recently, using this method we succeeded in revealing the presence of a magnon supercurrent by the analysis of an enhanced decay of the magnon BEC in a heated spot created by a probing laser beam \cite{Bozhko2016}. The main idea underlying the direct observation of the motion of the magnon BEC is to separate the area of supercurrent formation from the area in which it was observed. This enables the detection of temporal changes in the spatial distribution of the magnon density caused by the magnon supercurrent. Two independent light sources are used, one for the heating and the other for the BLS probing of the magnon gas as is shown in Fig.\,\ref{Fig1_Setup}a. A powerful blue laser light, which is completely absorbed by a 5.6-$\mu$m-thick YIG film, is used for the film heating. The green light of low power, which is able to penetrate the YIG film with moderate attenuation, is used for the probing. Both the blue laser source and the YIG sample are mounted together on a movable stage to ensure that the predefined position of the heated area on YIG-film surface remains unchanged. The stage motion relative to the focal green spot allows for the probing of the magnon gas density in different points of the sample.

\begin{figure}[t!]
	\includegraphics[width=1.\columnwidth]{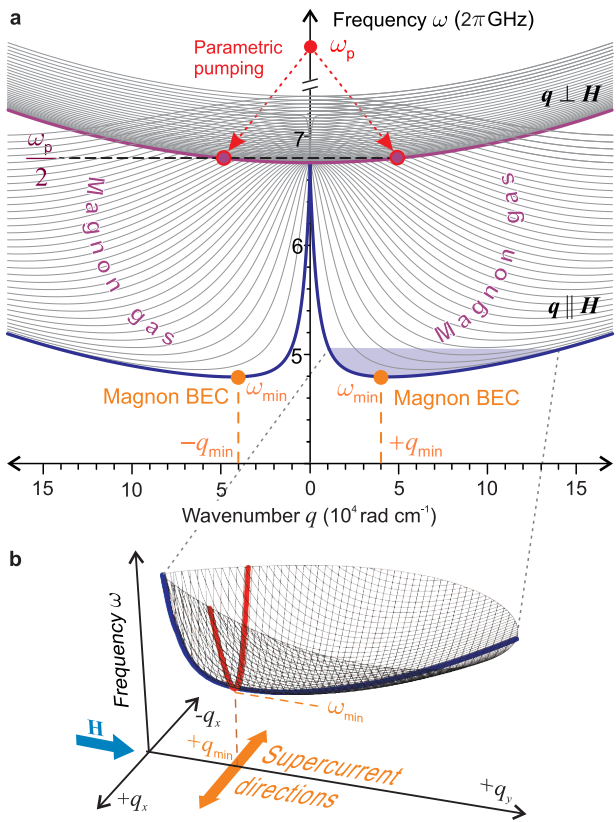}
	\caption{\label{Fig2_Spectra} \textbf{Magnon spectrum.} 
		\textbf{a}, Magnon spectrum of a 5.6-$\mu$m-thick YIG film magnetized in plane by a bias magnetic field $H=1690$\,Oe is presented for the wavevectors $\textbf{\textit{q}}$ perpendicular ($\pm  \textbf{\textit{q}} \perp \textbf{\textit{H}}$) and parallel ($\pm  \textbf{\textit{q}} \parallel \textbf{\textit{H}}$) to the applied field. For both wavevector orientations the first 35 thickness modes are shown. The red arrows illustrate the process of injection of magnon pairs at $\omega_\mathrm{p}/2$ frequency by means of parallel parametric pumping. Thermalization of the parametrically overpopulated magnon gas leads to the formation of Bose-Einstein condensates in two symmetric minima of the frequency spectrum $\omega(\textbf{\textit{q}})$ at $\textbf{\textit{q}} =\pm \textbf{\textit{q}}_\mathrm{min}$ (with $\textbf{\textit{q}}_\mathrm{min} \parallel \textbf{\textit{H}}$). Inset \textbf{b} shows the 3-dimensional view of a bottom part of the spectrum (the lowest fundamental mode $n=0$ within the shaded magenta area in panel \textbf{a}) calculated in the range up to 200\,MHz above the $\omega_\mathrm{min}$. The red curve shows the spectrum of magnons with $\pm \textbf{\textit{q}} \perp \textbf{\textit{H}}$ and the blue curve relates to the magnons with $+\textbf{\textit{q}} \parallel \textbf{\textit{H}}$. Due to much higher (about 21 times) steepness of the red curve near $(\omega_\mathrm{min}, \textbf{\textit{q}}_\mathrm{min})$, and, thus, proportionally smaller effective masses of the corresponding magnons, the magnon supercurrents predominantly propagate in a film plane across $\textbf{\textit{H}}$ and along the $\pm \textbf{\textit{q}}_x$ directions. \looseness=-1} 
\end{figure}

In order to achieve Bose-Einstein condensation \cite{Demokritov2006, Serga2014}, magnons are injected into the spin system of the YIG film via parallel parametric pumping \cite{Gurevich1996}, which is currently considered the most efficient technique for magnon excitation over a large wave-vector range. The process can be described by the splitting of a photon of a pumping electromagnetic wave with a wavevector of nearly zero and a pumping frequency $\omega_\mathrm{p}$ into two magnons with opposite wavevectors $\pm \textbf{\textit{q}}$ and a frequency of $\omega_\mathrm{p}/2$ (see Fig.\,\ref{Fig2_Spectra}a). The strength of the bias magnetic field $H_0=1690\,\mathrm{Oe}$ is chosen to allow for the injection of the magnon pairs slightly above the ferromagnetic resonance frequency $\omega_{_\mathrm{FMR}}$, where the parallel pumping process is most efficient \cite{Serga2012,Neumann2009}. The injected quasi-particles thermalize by way of the four-magnon scattering processes conserving both their number and the total energy \cite{Demidov2008, Clausen2015, Bozhko2015, Bozhko2017, Kreil2018}. Finally, when the total number of magnons reaches a threshold value, a magnon BEC forms \cite{Demokritov2006, Serga2014}. Due to the spatially confined microwave excitation, the BEC is formed in an oblong YIG region located just above the pumping resonator. This quasi-one-dimensional region extends for $50\,\mathrm{\mu m}$ along and for $1\,\mathrm{mm}$ across the direction of the bias magnetic field $\textbf{\textit{H}}$ (see Fig.\,\ref{Fig1_Setup}b). Due to the large anisotropy of the magnon spectrum (see Fig.\,\ref{Fig2_Spectra}b), the effective mass of the magnons, which is inversely proportional to the magnon dispersion coefficient $D(q_x,q_y)=\mathrm{d}^{\,2} \omega(\textbf{\textit{q}})/(2\mathrm{d} \textbf{\textit{q}}^2)$, strongly depends on the orientation of the magnon wavevector $\textbf{\textit{q}}$. The mass of the magnons with wavevectors placed perpendicular to the bias magnetic field is approximately $21$ times smaller than for magnons having wavevectors along the field \cite{Bozhko2016}. The magnon supercurrent is therefore expected to be approximately $21$ times stronger across the biased magnetic field and thus, along a large extent of the condensate. Taking into consideration these facts, we currently restrict ourselves only to one dimensional scanning, along the pumping resonator (Fig.\,\ref{Fig1_Setup}b).

\begin{figure*}[t]
	\includegraphics[width=\textwidth]{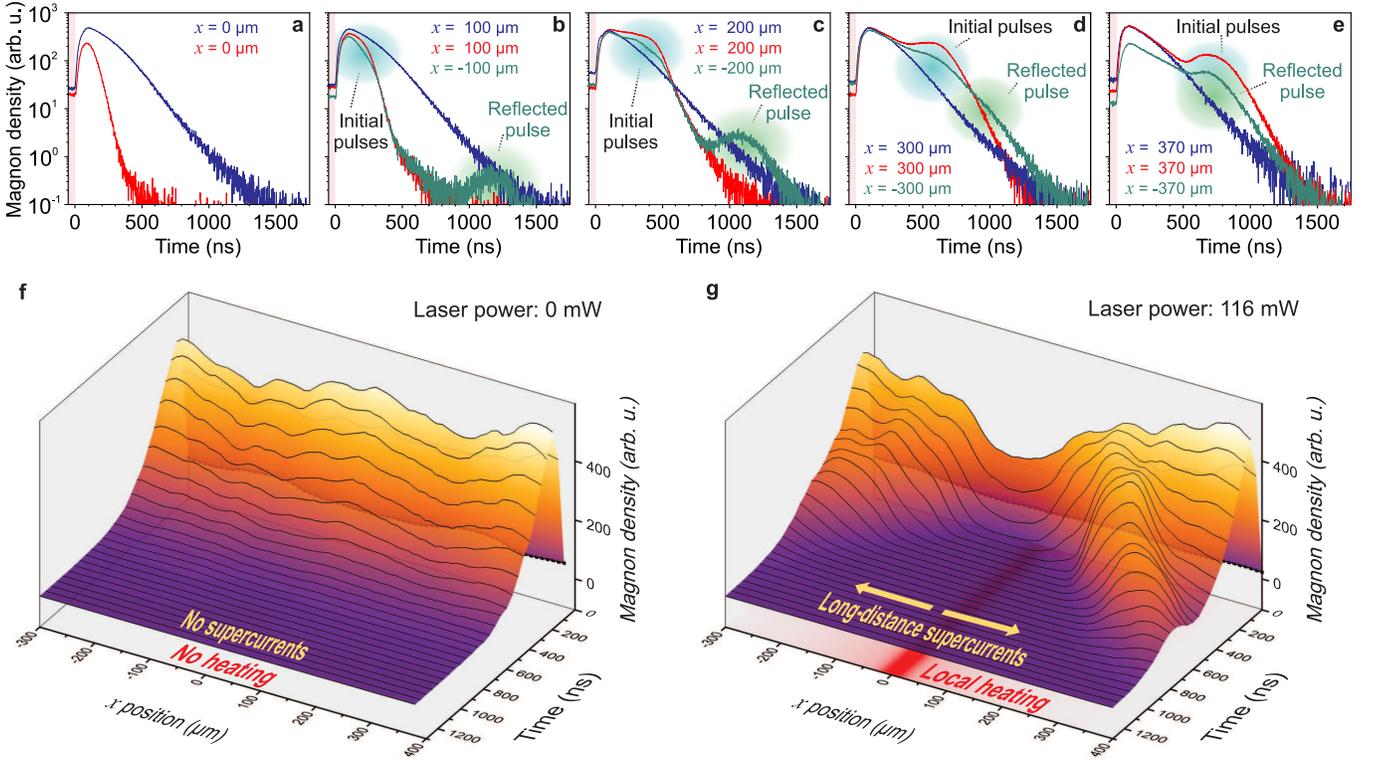}
	\caption{\label{Fig3_2-3Dcurrents} \textbf{Time and time-space population diagrams of the magnon Bose-Einstein condensate (BEC).} 
		\textbf{a}-\textbf{e}, Time evolutions of the temperature-uniform (blue curves) and locally heated (red and green curves, laser power is of 116\,mW) magnon BECs are shown in different spatial positions along the microstrip resonator. The vertical pink stripes mark the end part of the pumping pulse, which is switched off at zero moment of time. Formation of peaks of magnon density (marked by smeared blue ovals) and their propagation outwards the hot sample area are clearly visible on the background of a freely decaying BEC. The BLS pulses related to the density peak reflected from the left edge of the sample back to the hot area are visible on the green curves in panels \textbf{b}-\textbf{e} (marked by smeared green ovals). In panel \textbf{d}, the input and reflected peaks  on the green curve start to overlap in time. In panel \textbf{e}, these peaks on the green curve (measured just near the left sample edge) are perfectly overlapped.  \textbf{f}-\textbf{g}, Space-time evolutions of the magnon BEC after termination of the pumping pulse. After the  parametric pumping is switched off at zero moment of time, a BEC density peak is formed. \textbf{f}, Without external heating: the BEC density exponentially decays in time uniformly over space (cf. the blue curves shown in panels \textbf{a}-\textbf{e} in logarithmic scale). \textbf{g}, With external continuous heating by a blue laser: the heating does not change the spatial distribution of the magnon density in the course of pumping action (see the dotted curve at the zero moment of time). However, the forming magnon BEC is pushing out from the hot focal spot (the simulated temperature profile is shown in the plane below) and a magnon density dip surrounded by two magnon density humps appears in the heated area. These humps propagate in opposite directions along the microstrip and perpendicularly to the bias magnetic field $\textbf{\textit{H}}$ through the ``cold'' magnon BEC over distances of hundred micrometers. }
\end{figure*}

The results of the spatially-resolved measurements of the magnon density evolution at the bottom of the spin-wave spectrum are shown in Fig.\,\ref{Fig3_2-3Dcurrents} for different heating conditions. To understand the dynamics of the magnon BEC, we first establish how it behaves in a spatially uniform room-temperature profile. The reference measurement, performed without heating, shows that the spatial distribution of the magnon condensate along the microstrip resonator is uniform (see BEC time profiles shown by the blue curves in Figs.\,\ref{Fig3_2-3Dcurrents}a-e and a time-space diagram in Fig.\,\ref{Fig3_2-3Dcurrents}f). Similar to the results we presented in Ref.\,\cite{Serga2014}, the BLS signal, which is proportional to the magnon density, rises sharply after the microwave pumping pulse is switched off due to the intensification of the BEC formation process caused by so-called ``supercooling'' of the magnon gas \cite{Serga2014}. Due to the intrinsic magnon relaxation to the phonon bath, the density of the freely-evolving magnon BEC exponentially decreases and we observe a spatially uniform decay of the BLS signal.

We now focus on the temperature-gradient dependent behavior of the magnon condensate. The outcome of the experiment is strongly changed when an additional local heating by the blue laser light with power of 44\,mW, 81\,mW, or 116\,mW is applied. The red curves in Figs.\,\ref{Fig3_2-3Dcurrents}a-e and time-space diagrams in Fig.\,\ref{Fig3_2-3Dcurrents}g exemplify the BEC dynamics measured for the laser power of 116\,mW. Note, that the heating is applied continuously, thus the condensate forms in a predefined non-uniform temperature profile. In contrast to the uniform-temperature reference measurement, the magnon density distribution becomes spatially-inhomogeneous already during the formation of the condensate. A pronounced dip in the magnon population is formed in the heated region of the sample (Fig.\,\ref{Fig3_2-3Dcurrents}g). This reduction in the condensate population is closely related to the previously reported enhancement of the relaxation rate of magnon BEC in the hot spot \cite{Bozhko2016, Kreil2018} (cf. the relaxation rates of the blue and red curves in Fig.\,\ref{Fig3_2-3Dcurrents}a) and is associated with the formation of a magnon supercurrent, flowing out of the high temperature region. 

Furthermore, after some time we start to observe the formation of two magnon density peaks, traveling outwards from the hot spot. It is necessary to emphasize, that their propagation occurs in the sample area with a uniform  temperature distribution. As is shown in Fig.\,\ref{Fig3_2-3Dcurrents}g, the heating by the blue laser is rather local and does not extend beyond  $\pm 200\,\mathrm{\mu m}$ from the center of the focal spot. At the same time, the peak visible in Fig.\,\ref{Fig3_2-3Dcurrents}g, propagates to more than $400\,\mathrm{\mu m}$. The maximum propagation distance, observed in our experiment, is $600\,\mathrm{\mu m}$, being restricted by the right edge of the YIG sample. Moreover, one of the density peaks is efficiently reflected from the nearby YIG-film edge on the left side of the sample (see Fig.\,\ref{Fig2_Spectra}b) and propagates hundreds of microns back towards the hot spot (see the second peaks on green curves in Fig.\,\ref{Fig3_2-3Dcurrents}b-d). Such a  dynamic is common for all used heating powers. Stronger laser power leads to a faster and deeper appearance of the density dips. This results into the formation of more intense BEC humps but does not visibly change the propagation speed of these humps, which is approximately 350\,$\mathrm{m}\,\mathrm{s}^{-1}$. The observed distant propagation of magnon density peaks through the uniform BEC cannot be understood using the model assuming the accumulation of a BEC phase in a thermally created magnetic potential used in Ref.\,\cite{Bozhko2016} and a subsequent supercurrent.  

\bigskip
{
	\fontsize{11pt}{10.0pt}
	\selectfont
	\noindent\textcolor{black}{\textbf{Second sound and time-evolution of inhomogeneous magnon BEC}}}

\noindent To discuss the observed time-evolution of the total number of magnons near the lower end of the frequency spectrum $N_\mathrm{tot}= N_\mathrm{b} + N_\mathrm{c}$, which comprises the gaseous bottom magnons $N_\mathrm{b}$ and condensed magnons $N_\mathrm{c}$, we need to clarify the initial conditions for $N_\mathrm{b}$ and $N_\mathrm{c}$ at the moment of time $t  = +  0$, immediately after the termination of pumping power. The previous wavevector- and time-resolved BLS studies in Refs.\,\cite{Serga2014} revealed that strong electromagnetic pumping overheats the magnon gas and can thus prevent the condensation process, which develops rapidly after the pumping is switched off \cite{Serga2014,Bozhko2015,Clausen2015,Demidov2008}. In addition, the experimentally observed magnon density is not affected by the local heating during  pumping (see the dashed line at $t = + 0$ in Fig.\,\ref{Fig3_2-3Dcurrents}). It indicates that in the entire pumping area no BEC exists under given experimental conditions before the pumping is switched off. Thus, we can assume that for $t\leqslant 0$ $N_\mathrm{tot} \approx N_\mathrm{b}$ and $N_\mathrm{c}\ll N_\mathrm{b}$ (or equal to zero). 

Our understanding of what happen for $t \geq 0$ in the presence of the hot spot, where the locally reduced saturation magnetization creates a frequency well, is at an infancy state. An essential progress in this direction requires purposeful experimental study, which will than serve as a basis for a microscopic analytical theory. For the time being the experiments can be reasonably well interpreted under a conjecture that fast thermalization of $N_\mathrm{b}$ for $t > 0$ creates a high-intensity condensate mainly (or only) in the hot spot. To find possible reasons for such localization we can speculate that the almost-space-homogeneous chemical potential touches the bottom of the frequency spectrum initially at the hot spot. If so, the magnon BECs with $\pm q_\textrm{min}$ (see Fig.\,\ref{Fig2_Spectra}a) will try to escape from this area due to their nonlinear repulsion \cite{Dzyapko2017}. As is evident in our analysis of the propagation data (see below), the repulsive frequency shift $\Omega_\mathrm{NL}$ is about one order of magnitude larger that the depth of the frequency well $\delta \omega(T)$, induced by the heating (see Methods). Therefore, the magnon BEC can successfully escape the well, creating propagating second sound pulses. These pulses can be formally expanded into Fourier series of monochromatic waves, which propagate almost with the same velocity due to the linearity of the dispersion law (3a) in the small wavevector limit. This explains why the observed density humps propagate practically without changing their form. Moreover, the travelling monochromatic waves should reflect simultaneously from the sample edge, creating a hump of the same form, propagating in the opposite direction. This is exactly what we observe as a reflected pulse on the green curves in Fig.\,\ref{Fig3_2-3Dcurrents}e.

We should notice that the massive transition of the gaseous bottom magnons to the effluent magnon condensate should significantly reduce their concentration $N_\mathrm{b}$ in the center of the hot spot and some diffusion of these quasiparticles towards the hot spot to replace the escaping BEC magnons is expected. One can guess that the effective diffusion coefficient is given by the dispersion coefficient $D_x$. If this is so, the characteristic speed of the diffusive broadening of the magnon density hole of width $\Delta$ will be approximately $D_x\Delta \approx 17\,\mathrm{m}\,\mathrm{s}^{-1}$ for the maximum laser power $P=116$\,mW (see Tab.\,\ref{t:1}). This is much smaller than the experimentally determined speed ($\approx 350\,\mathrm{m}\,\mathrm{s}^{-1}$) of the propagating BEC humps. Therefore, the diffusion of the normal component can be neglected and the dip in the total magnon density $N_\mathrm{tot}$ should appear as is indeed observed in our experiment (see Fig.\,\ref{Fig3_2-3Dcurrents}g).

The problem of propagation of the magnon density humps outside of the hot spot can be solved by analysis of the dynamic behaviour of the magnon BEC confined in a long but narrow area along the microstrip resonator (Fig.\,\ref{Fig1_Setup}b) using a one-dimensional version of the Gross-Pitaevskii equation (GPE): 
\begin{eqnarray}\label{GPE}
i \frac{\partial C}{\partial t} =-D_x \frac{\partial^2 C}{\partial x^2} + W |C|^2 C \, .
\end{eqnarray}
Here $C(x,t)$ is the BEC wave function, $D_x=\partial^{\,2} \omega(\textbf{\textit{q}})/(2\partial q_x^2)$ is the dispersion coefficient at the minimum frequency, and $W$ is the effective amplitude of the four-wave interaction, responsible for the nonlinear frequency shift. The first term in the right-hand side of Eq.\,\eqref{GPE} is the density of kinetic energy, the second one is the density of potential energy in mean field approximation. In our case $W$ is positive, corresponding to the repulsive interaction in the system of two interacting BECs \cite{Dzyapko2017}.

The stationary solution of GPE\,\eqref{GPE} has the form
\begin{eqnarray}\label{stat}
C_0(x,t)= \sqrt {N_\mathrm{c}} \exp (- i \Omega\Sb{NL}t)\,, \quad \Omega\Sb{NL}= WN_\mathrm{c}\, ,
\end{eqnarray}
where $|C_0|^2=N_\mathrm{c}$ is the density of magnons in the BEC state.
The dispersion law for the small perturbation $c(x,t)\propto \exp i [k x -\Omega(k)t]$ of the stationary solution\,\eqref{stat} found by Bogoliubov has the form \cite{Pitaevskii2003}:
	\begin{eqnarray}\label{3}
    \Omega(k) = k \sqrt {2D_x \Omega\Sb{NL} ({1+ D_x k^2/2 \Omega\Sb{NL}})} \ ,
	\end{eqnarray}
where $\Omega$ and $k$ are the angular frequency and the wave number of the perturbation, respectively. In the long-wavelength limit $D_x k^2 \ll 2 \Omega\Sb{NL}$ this perturbation has the linear dispersion law
	\begin{eqnarray}\label{4}
    \Omega(k) = c_\textrm{s} k \, , \quad c\sb s= \sqrt {2D_x \Omega\Sb{NL}} \, , 
    \end{eqnarray}
and may be understood as a second sound propagating through the BEC with the velocity $c\sb s$. 
It is important to note that according to Landau's criterion $\Omega_k > c_s k$ \cite{Pitaevskii2003}, a BEC with repulsive interaction, as realised in our case, is an inviscid superfluid.
	
At the same time, for large $k$ the BEC contribution may be neglected and the standard quadratic dispersion law applies near the minimum of the spectra: $\Omega(k) =  D_x k^2$.

In order to check to which of these two limiting cases our experimental situation is closer, we measured the width $\Delta$ of the magnon density peaks at half of their height for different heating powers. These measurements enabled us to estimate the characteristic value of the second sound wavenumbers $k\simeq \pi/\Delta$ (the first spatial Fourier harmonic), which give the main contributions to the propagating pulse, see Tab.\,\ref{t:1}. Additionally, for each laser power used for heating, we measured the position $x(t)$ of the pulse maximum as a function of the propagation time $\tau=t-t_0$, where $t_0$ is some initial moment of time, for which the density peak is already well shaped and such analysis is therefore possible. In order to reveal a possible dependence of the sound velocity on the propagation time we fitted the experimental data by a second order polynomial function $x(t)= c\sb s \tau - \delta c\sb s \tau^2/ 2 \tau\sb{max}$. Here $\tau\sb{max}$ is the maximum propagation time, limited by the length of our sample. Using these fits we can estimate the initial velocity of the propagating peak $c\sb s$ (at $t=t_0$, $\tau=0$) and its final velocity $c\sb s - \delta c\sb s$ at $\tau=\tau\sb{max}$. The results are given in Tab.\,\ref{t:1}.

\begin{table}
	\centering
	\begin{tabular}{|c|c|c|c|c|}
		\hline
		$P_\mathrm{L}\,$ & $\Delta$ & $k\,$           & $c\sb s(t_0)\,$ & $c\sb s (t\sb {max})\, $ \cr
		mW               & $\mu$m   & rad cm$^{-1}$   & m s$^{-1}$      & m s$^{-1}$     \cr \hline
		116              & 44       & 714             & 325             & 306            \cr
		81               & 72       & 436             & 361             & 220            \cr
		44               & 88       & 357             & 394             & 133            \cr
		\hline
	\end{tabular}
	\caption{\label{t:1} The parameters of the propagating pulses at different heating laser powers $P_\mathrm{L}$: $\Delta$ is the full width of the pulse at half of their height, $k=\pi/\Delta$ is the characteristic wavenumber of the second sound that mainly contribute to the pulse profile, $c\sb s (t_0)$ and $c\sb s (t\sb{max})$ is the initial and final speeds of the propagating pulses.  }
\end{table}

Our main finding in Tab.\,\ref{t:1} is that the initial velocity  $c\sb s(t_0)$ of the peak only weakly depends on its wavenumber $k$: it varies by about 20\% as $k$ changes twice. A natural explanation of this fact is based on the assumption that the propagating peak consists of the second sound waves over the background of the magnon BEC with a linear dispersion law\,\eqref{4}. If so, the velocity $c\sb s(t_0)$ should indeed be $k$-independent. To check whether the long-wavelength limit assumption is really valid in our case, we estimate the value of the nonlinear frequency shift $\Omega\Sb{NL}=W N_0$ \eqref{stat} and the product $D_x k^2/2$ by taking $c\sb s=325 $\,m\,s$^{-1}$ and $k\approx 714$\,rad\,cm$^{-1}$ from Tab.\,\ref{t:1} for $P_\mathrm{L}=116\,$mW and calculating $D_x\approx 7.4\, \mathrm{cm}^2\mathrm{s}^{-1}\mathrm{rad}^{-1}$.
The resulting estimates are $\Omega\Sb{NL} \approx 2 \pi \cdot 11.3 \,\mathrm{MHz}$ and $D_x k^2/2 \approx 2 \pi \cdot 0.3\,\mathrm{MHz}$.
It is evident, that the long-wavelength limit is well satisfied, supporting the suggested second sound scenario.

The second argument in favour of the suggested second sound scenario for the propagating pulses is the time dependence of their velocity. During pulse propagation, the amplitude of the background condensate decays and this consequently leads to a decrease in the sound velocity, as indeed  observed in the experiment (see Tab.\,\ref{t:1}). 

Using the estimated value of $\Omega\Sb{NL}$, we can also estimate the coherence length, which determines the distance at which the linear dispersion law \eqref{4} is changed to the quadratic one. The resulting estimate $\xi = \sqrt {D_x / \Omega_\mathrm{NL}}\simeq 1.2\, \mu\mathrm{m}$ is much smaller than the total propagation length $L\simeq 600 \,\mu$m and the peak width $\Delta\sim 44 \,\mu$m. 
On the other hand, the coherent length determines the sizes of possible topological singularities in the magnon BEC. For example, the diameter of vortex cores in the magnon condensate measured under similar experimental conditions in Ref.\,\cite{Nowik-Boltyk2012} is about $1\,\mu$m which is comparable with our estimate.

\bigskip
{
	\fontsize{11pt}{10.0pt}
	\selectfont
	\noindent\textcolor{black}{\textbf{Discussion}}}

\noindent 
The results presented in this article, addressing the spatially-resolved probing of the dynamics of a magnon BEC, provide direct evidence of supercurrent-related motion of the condensate outwards from the heated spot. The observed occurrence of the magnon BEC propagation outside of the temperature gradient is associated with the excitation of a new type of the second sound: magnon second sound in the magnon BEC condensate. The newly discovered second sound differs from the second sound in dielectrics \cite{Gurevich1988,Ackerman1966,McNelly1970,Lee2015}, in which the phonons can be described  in terms of their occupation numbers only, not taking into account their phases. It also differs from the second sound in superfluid He and BEC in diluted atomic systems, where the wave function describes the distribution of real atoms and not of quasiparticles, as in our case. 

The suggested propagation scenario requires further detailed experimental and theoretical investigations. For example, one needs to account for the interactions of the gaseous magnons with the BEC magnons, by formulating a  model, analogous to the two-fluid model of superfluid He \cite{Landau1941}. Having said that, we should emphasis that from a practical point of view, the three observed phenomena: (i) the transition from the supercurrent-type to the second-sound-type propagation regimes, (ii) the excitation of the second-sound-pulses, and (iii) the possibility of a long-distance spin-transport in the magnon BEC, has already paved a way for the application of the magnon macroscopic quantum states for low-loss data transfer and information processing in perspective magnon spintronic devices \cite{Nakata2014,Chumak2015,Schneider2018}.

\bigskip
{
	\fontsize{11pt}{10.0pt}
	\selectfont
	\noindent\textcolor{black}{\textbf{Acknowledgements}}}

\noindent Financial support of the Deutsche Forschungsgemeinschaft through the Collaborative Research Center SFB/TRR-49 ``Condensed Matter Systems with Variable Many-Body Interactions'' (project INST 161/544-3) as well as financial support of the European Research Council within the Advanced Grant 694709 SuperMagnonics -- ``Supercurrents of Magnon Condensates for Advanced Magnonics'' is gratefully acknowledged.

\smallskip

\bigskip


{\fontfamily{phv}
	\fontsize{10pt}{10.0pt}
	\selectfont
	\noindent\textcolor{black}{\textbf{METHODS}}}

\smallskip
{\fontsize{9pt}{9.0pt}
	\selectfont 
	
\noindent\textbf{Sample.}
The Yttrium Iron Garnet (YIG, $\mathrm{Y_{3}Fe_{5}O_{12}}$) \cite{Cherepanov1993} sample is 5\,mm long and 1\,mm wide. The single-crystal YIG film \cite{Glass1988} of 5.6\,$\mu$m thickness has been grown in the (111) crystallographic plane on a Gadolinium Gallium Garnet (GGG, $\mathrm{Gd_{3}Ga_{5}O_{12}}$) substrate by liquid-phase epitaxy at a Department of Crystal Physics and Technology of the Scientific Research Company ``Carat'' (Lviv, Ukraine).

\smallskip
\noindent\textbf{Experimental setup.} A sketch of the experimental setup is shown in Fig.~\ref{Fig1_Setup}a. It consists of microwave and optical parts. The microwave part includes a microwave generator, which is used as a source for pumping pulses (pulse duration $2\,\mathrm{\mu s}$, repetition time $1\,\mathrm{ms}$, carrier frequency $13.6\,\mathrm{GHz}$) followed by a power amplifier, which drives a microstrip resonator circuit with a peak power of $12.6\,\mathrm{W}$. The $50\,\mathrm{\mu m}$ wide half-wavelength microstrip resonator fabricated on top of an alumina substrate is used to further increase the amplitude of the pumping microwave magnetic field and its spatial localization. The YIG sample is positioned on top of the middle part of the resonator, in the area of maximal microwave magnetic field. 

\noindent The optical part (Fig.~\ref{Fig1_Setup}a) is used both for the probing of the magnon Bose-Einstein condensate (BEC) by means of Brillouin light scattering spectroscopy and for the controlled local heating of the YIG sample. Its main parts are the probing green laser (single-mode, $532\,\mathrm{nm}$ wavelength), an acousto-optic modulator (AOM), heating blue laser (multi-mode, $405\,\mathrm{nm}$ wavelength), and a tandem Fabry-P\'erot interferometer (TFPI). 

\noindent As we have shown in our previous work \cite{Bozhko2016}, an excessive heating of the sample by the probing laser can lead to the formation of a magnon supercurrent. In order to minimize the influence of the probing beam on the magnon dynamics we utilize an acousto-optic modulator, which is used for chopping the probing beam into pulses to reduce a parasitic heating of the sample. The pulsed probing beam (pulse duration $6\,\mathrm{\mu s}$, peak power $9\,\mathrm{mW}$) is then focused onto the sample surface into a focal spot of $20\,\mathrm{\mu m}$ in diameter. The scattered light is directed to a multipass tandem Fabry-P\'{e}rot interferometer \cite{Sandercock1981,Mock1987,Hillebrands1999} for frequency selection with resolution of 100\,MHz. At the output of the interferometer a single photon counting avalanche diode detector is placed. The output of the detector is connected to a counter module synchronized with a sequence of microwave pulses. Every time the detector registers a photon, this event is recorded to a database which collects the number of arrived photons ensuring such a time resolution of 250\,ps. The frequency of the interferometer transmission is also recorded, thus providing frequency information for each detected photon. \looseness=-1

\smallskip
\noindent\textbf{Frequency- and wavevector-resolved Brillouin light scattering spectroscopy.} 
Brillouin light scattering can be understood as the diffraction of the probing light from a moving Bragg grating created by a magnon mode. Some portion of the scattered light, which is proportional to the number of magnons in this mode, is shifted in frequency by the frequency of the mode. In addition, the diffraction from the grating leads to a transfer of momentum during this process: the in-plane component of the wavevector $\textbf{\textit{q}}_\mathrm{L}$ of the incident light is inverted by a magnon mode if the magnon wavenumber $q$ satisfies the Bragg condition $q = -2q_\mathrm{L} \sin(\Theta)$, where $\Theta$ is the angle of incidence. Simultaneously, the out-plane wavevector of the probing light is inverted due to its reflection from the metal microstrip underlying a semi-transparent YIG sample. By setting of the angle $\Theta$ equal to $9.7^\circ$, selection of in-plane magnons with wavenumbers $q_y \approx 4\cdot10^4\,$rad/cm around one of the minima of magnon spectra (Fig.~\ref{Fig2_Spectra}) is implemented in our setup.\cite{Sandweg2010} The wavevector resolution was $\pm 2\cdot10^3\,\textrm{rad/cm}$ allowing for a rather selective observation of magnon dynamics around the magnon energy minimum at $q_y = q_\textrm{min}$. It allows to avoid in our transport measurements a possible spurious contribution of travelling magnon-phonon hybrid quasiparticles \cite{Bozhko2017}.

\smallskip
\noindent\textbf{Optical heating of the sample.} In order to locally heat the sample in a spatial point separated from the probing spot an additional continuous $405\,\mathrm{nm}$ wavelength laser is used. The reasoning behind the choice of this laser wavelength is twofold. Firstly, it allows an efficient heating of the thin YIG film, since the absorption of the light in the YIG layer is inversely proportional to the wavelength of the light. Secondly, the chosen wavelength is well separated from the probing laser wavelength, and therefore rather easily filtered out to exclude any possible influence on the detection system. The heating beam focused into a focal spot of $80\,\mathrm{\mu m}$ in diameter provides a local increase of the sample temperature by $40\,\mathrm{K}$ at maximal laser power of $116\,\mathrm{mW}$. Spatial-resolved probing of the magnon dynamics is performed by a controlled displacement of the sample using a precise linear positioning stage. The whole stage is placed directly between the poles of the electromagnet, ensuring high field uniformity and stability. Since the heating laser is located on the same stage, the position of the heating spot is fixed relative to the sample and to the excitation circuit (and, thus, relative to the created magnon BEC). Thus, the described setup allows for space-resolved measurements of the magnon dynamics across the heated and cold areas of the sample as it is shown in Fig.\,\ref{Fig1_Setup}. The spatial resolution (scanning step) is set to $10\,\mathrm{\mu m}$, which corresponds to half the probing focal spot size.

\smallskip
\noindent\textbf{Temperature and BEC frequency shift in the hot spot.} The temperature profiles of the heated sample were determined by solving a 3D heat-transfer model of the experimental setup using the \textit{COMSOL Multiphysics} software \cite{COMSOL} as it was done in Ref.\,\cite{Bozhko2016}. Hereby, the conventional heat conduction differential equation is solved under consideration of the boundary conditions applied to the model, the material parameters of the used materials and the applied heat source. Heat is deposited exponentially along the film thickness and has a Gaussian distribution in the film plane reflecting the shape of the laser focal spot. The calculated temperature difference $\Delta T$ between the centre of the laser focal spot and the cold film for the maximal laser power $P_\mathrm{L}$ of 116\,mW is about 40\,K. The corresponding frequency difference $\delta \omega (T) $ caused by the temperature induced decrease in the saturation magnetization of an YIG film \cite{Cherepanov1993,Bozhko2016,Mihalceanu2018} is about $2 \pi \cdot 4.6\,\mathrm{MHz}$. This value is much smaller than the nonlinear frequency shift $\Omega\Sb{NL}$ in our experiment.

\smallskip

}



\begin{thebibliography}{99}

	{\fontfamily{phv}
	\fontsize{8pt}{9.0pt}
	\selectfont


\bibitem{Einstein1924} Einstein, A.
Quantentheorie des einatomigen idealen Gases.
\textit{Sitz.ber. Preuss. Akad. Wiss. Phys.} \textbf{22}, 261--267 (1924).

\bibitem{Einstein1925} Einstein, A.
Quantentheorie des einatomigen idealen Gases Zweite Abhandlung.
\textit{Sitz.ber. Preuss. Akad. Wiss. Phys.} \textbf{23}, 3--14 (1925).

\bibitem{Borovik-Romanov1984} Borovik-Romanov, A.S., Bun'kov, Yu.M., Dmitriev, V.V. \& Mukharskii, Yu.M.
Long-lived induction signal in superfluid $^3$He-B.
\textit{JETP Lett.} \textbf{40}, 1033--1037 (1984).

\bibitem{Anderson1995} Anderson, M.H., Ensher, J.R., Matthews, M.R., Wieman, C.E. \& Cirnell, E.A.
Observation of Bose-Einstein condensation in a dilute atomic vapor.
\href{https://doi.org/10.1126/science.269.5221.198}{\textit{Science} \textbf{269}, 198--201 (1995).}

\bibitem{Davis1995} Davis, K.B. \textit{et al.}
Bose-Einstein condensation in a gas of sodium atoms.
\href{https://doi.org/10.1103/PhysRevLett.75.3969}{\textit{Phys. Rev. Lett.} \textbf{75}, 3969--3973 (1995).}

\bibitem{Butov2001} Butov, L.V. \textit{et al.},
Stimulated scattering of indirect excitons in coupled quantum wells: signature of a degenerate Bose-gas of excitons.
\href{https://doi.org/10.1103/PhysRevLett.86.5608}{\textit{Phys. Rev. Lett.} \textbf{86}, 5608--5611 (2001).}

\bibitem{Kasprzak2006} Kasprzak, J. \textit{et al.},
Bose-Einstein condensation of exciton polaritons.
\href{https://doi.org/10.1038/nature05131}{\textit{Nature} \textbf{443}, 409--414 (2006).}


\bibitem{Safonov1989} Kalafati, Yu.D., \& Safonov, V.L.
Possibility of Bose condensation of magnons excited by incoherent pump.
\href{http://www.jetpletters.ac.ru/ps/1126/article_17065.shtml}{\textit{JETP Lett.} \textbf{50}, 149 (1989).}


\bibitem{Demokritov2006} 
Demokritov, S.O. \textit{et al.}
Bose-Einstein condensation of quasi-equilibrium magnons at room temperature under pumping.
\href{https://doi.org/10.1038/nature05117}{\textit{Nature} \textbf{443}, 430--433 (2006).}

\bibitem{Balili2007} Balili, R., Hartwell, V., Snoke, D., Pfeiffer, L. \& West, K.
Bose-Einstein condensation of microcavity polaritons in a trap.
\href{https://doi.org/10.1126/science.1140990}{\textit{Science} \textbf{316}, 1007--1010 (2007).}

\bibitem{Klaers2010} Klaers, J., Schmitt, J., Vewinger, F. \& Weitz, M.
Bose-Einstein condensation of photons in an optical microcavity.
\href{https://doi.org/10.1038/nature09567}{\textit{Nature} \textbf{468}, 545--548 (2010).}

\bibitem{Bardeen1957} Bardeen, J., Cooper, L.N. \& Schrieffer, J.R.
Microscopic theory of superconductivity.
\href{https://doi.org/10.1103/PhysRev.106.162}{\textit{Phys. Rev.} \textbf{106}, 162--164 (1957).}

\bibitem{Kapitza1938} Kapitza, P.
Viscosity of Liquid Helium Below the $\lambda$-Point. 
\href{https://doi.org/10.1038/141074a0}{\textit{Nature} \textbf{141} 3558 (1938).}

\bibitem{Allen1938} Allen, J.F. \&  Misener, A.D. 
Flow of Liquid Helium II.
\href{https://doi.org/10.1038/142643a0}{\textit{Nature} \textbf{142} 3597 (1938).}

\bibitem{Landau1941} Landau, L.D. 
Theory of the Superfluidity of Helium II. 
\href{https://doi.org/10.1103/PhysRev.60.356}{\textit{Phys. Rev.} \textbf{60}, 356--358 (1941).}

\bibitem{Osheroff1972} Osheroff, D.D., Richardson, R.C. \& Lee, D.M.
Evidence for a new phase of solid He$^3$
\href{https://doi.org/10.1103/PhysRevLett.28.885}{\textit{Phys. Rev. Lett.} \textbf{28}, 885 (1972).}

\bibitem{Leggett2004} Leggett, A.J.
Nobel lecture: superfluid $^3$He: the early days as seen by the theorist.
\href{https://doi.org/10.1103/RevModPhys.76.999}{\textit{Rev. Mod. Phys.} \textbf{76}, 999 (2004).}

\bibitem{Page2017} Page, D., Lattimer, J.M., Prakash, M. \& Steiner, A.W.
in \href{https://doi.org/10.1093/acprof:oso/9780198719267.001.0001}{\textit{Novel Superfluids} Vol. 2 (eds Bennemann, K.H. \& Ketterson, J.B.)}
505-579 (Oxford Univ. Press, 2014).


\bibitem{Bunkov2018} Bunkov, Yu.M. \& Safonov, V.L.
Magnon condensation and spin superfluidity.
\href{https://doi.org/10.1016/j.jmmm.2017.12.029}{\textit{J. Magn. Magn. Mater.} \textbf{452}, 30 (2018).}


\bibitem{Sonin2010} Sonin, E.B.
Spin currents and spin superfluidity.
\href{https://doi.org/10.1080/00018731003739943}{\textit{Advances in Physics} \textbf{59}, 181--255 (2010).}

\bibitem{Matthews1999} Matthews, M.R. \textit{et al.},
Vortices in a Bose-Einstein condensate.
\href{https://doi.org/10.1103/PhysRevLett.83.2498}{\textit{Phys. Rev. Lett.} \textbf{83}, 2498--2501 (1999).}

\bibitem{Raman1999} Raman, C. \textit{et al.},
Evidence for a critical velocity in a Bose-Einstein condensed gas.
\href{https://doi.org/10.1103/PhysRevLett.83.2502}{\textit{Phys. Rev. Lett.} \textbf{83}, 2502--2505 (1999).}

\bibitem{Borovik-Romanov1988} Borovik-Romanov, A.S. \textit{et al.} 
Observation of a spin-current analog of the Josephson effect.
\textit{JETP Lett.} \textbf{47}, 478--482 (1988).

\bibitem{Bunkov2009} Bunkov, Yu.M.
Spin superfluidity and coherent spin precession (Fritz London Memorial Prize Lecture).
\href{http://dx.doi.org/10.1088/0953-8984/21/16/164201}{\textit{J. Phys. Condens. Matter} \textbf{21}, 164201 (2009).}

\bibitem{Bunkov2012} Bunkov, Yu.M. \textit{et al.}
High-T$_\mathrm{c}$ spin superfluidity in antiferromagnets.
\href{https://doi.org/10.1103/PhysRevLett.108.177002}{\textit{Phys. Rev. Lett.} \textbf{108}, 177002 (2012).}

\bibitem{Amo2009} Amo, A. \textit{et al.}
Superfluidity of polaritons in semiconductor microcavities.
\href{https://doi.org/10.1038/nphys1364}{\textit{Nature Phys.} \textbf{5}, 805--810 (2009).}

\bibitem{Bozhko2016} 
Bozhko, D.A. \textit{et al.}
Supercurrent in a room-temperature Bose-Einstein magnon condensate.
\href{https://doi.org/10.1038/nphys3838}{\textit{Nature Phys.} \textbf{12}, 1027 (2016).}

\bibitem{Bunkov2008} Bunkov, Yu.M. \& Volovik, G.E.
Spin vortex in magnon BEC of superfluid $^3$He-B.
\href{https://doi.org/10.1016/j.physc.2007.11.026}{\textit{Physica C} \textbf{468}, 609--612 (2008).}

\bibitem{Nowik-Boltyk2012}
Nowik-Boltyk, P., Dzyapko, O., Demidov, V.E., Berloff, N.G. \& Demokritov, S.O. 
Spatially non-uniform ground state and quantized vortices in a two-component Bose-Einstein condensate of magnons. 
\textit{Sci. Rep.} \textbf{2}, 1 (2012).


\bibitem{Pitaevskii1968} Pitaevskii, L.P. 
Second sound in solids.
\href{http://iopscience.iop.org/article/10.1070/PU1968v011n03ABEH003839/pdf}{\textit{Usp. Fiz. Nauk} \textbf{95}, 139--144 (1968).}

\bibitem{Gurevich1988} Gurevich, V.L. 
Second sound in ferroelectrics.
\href{http://www.jetp.ac.ru/cgi-bin/dn/e_067_01_0206.pdf}{\textit{JETP} \textbf{67}, 1, 206 (1988).}

\bibitem{Ackerman1966} Ackerman, C.C., Bertman, B., Fairbank, H.A. \& Guyer, R.A. 
Second sound in solid helium.
\href{https://doi.org/10.1103/PhysRevLett.16.789}{\textit{Phys. Rev. Lett.} \textbf{16}, 789 (1966).}

\bibitem{McNelly1970} McNelly, T.F. \textit{et al.} 
Heat pulses in NaF: onset of second sound.
\href{https://doi.org/10.1103/PhysRevLett.24.100}{\textit{Phys. Rev. Lett.} \textbf{24}, 100 (1970).}

\bibitem{Lee2015} Lee, S., Broido, D., Esfarjani, K. \& Chen, G.
Hydrodynamic phonon transport in suspended graphene.
\href{https://doi.org/10.1038/ncomms7290}{\textit{Nat. Comm.} \textbf{6}, 6290 (2015).}


\bibitem{Gurevich1996} 
Gurevich, A.G. \& Melkov, G.A.
\textit{Magnetization Oscillations and Waves} (CRC Press, New York, 1996).

\bibitem{Cherepanov1993} Cherepanov, V., Kolokolov, I. \& L'vov, V.
The saga of YIG: spectra, thermodynamics, interaction and relaxation of magnons in a complex magnet.
\href{http://dx.doi.org/10.1016/0370-1573(93)90107-O}{\textit{Phys. Rep. -- Rev. Sec. Phys. Lett.} \textbf{229}, 81--144 (1993).}

\bibitem{Serga2010}	Serga, A.A., Chumak, A.V. \& Hillebrands, B. 
YIG magnonics. 
\href{http://dx.doi.org/10.1088/0022-3727/43/26/264002}{\textit{J. Phys. D: Appl. Phys.} \textbf{43}, 264002 (2010).}


\bibitem{Serga2014} 
Serga, A.A. \textit{et al.}
Bose-Einstein condensation in an ultra-hot gas of pumped magnons.
\href{https://doi.org/10.1038/ncomms4452}{\textit{Nat. Commun.} \textbf{5}, 3452 (2014).}

\bibitem{Serga2012} 
Serga, A.A. \textit{et al.}
Brillouin light scattering spectroscopy of parametrically excited dipole-exchange magnons.
\href{https://doi.org/10.1103/PhysRevB.86.134403}{\textit{Phys. Rev. B} \textbf{86}, 134403 (2012).}

\bibitem{Neumann2009} 
Neumann, T., Serga, A.A., Vasyuchka, V.I. \& Hillebrands, B. 
Field-induced transition from parallel to perpendicular parametric pumping for a microstrip transducer. 
\href{https://doi.org/10.1063/1.3130088}{\textit{Appl. Phys. Lett.} \textbf{94}, 192502 (2009).}

\bibitem{Demidov2008} 
Demidov, V.E. \textit{et al.} 
Magnon kinetics and Bose-Einstein condensation studied in phase space. 
\href{https://doi.org/10.1103/PhysRevLett.101.257201}{\textit{Phys. Rev. Lett.} \textbf{101}, 257201 (2008).}

\bibitem{Clausen2015} 
Clausen, P. \textit{et al.}
Stimulated thermalization of a parametrically driven magnon gas as a prerequisite for Bose-Einstein magnon condensation.
\href{https://doi.org/10.1103/PhysRevB.91.220402}{\textit{Phys. Rev. B} \textbf{91}, 220402(R) (2015).}

\bibitem{Bozhko2015} 
Bozhko, D.A. \textit{et al.}
Formation of Bose-Einstein magnon condensate via dipolar and exchange thermalization channels.
\href{https://doi.org/10.1063/1.4932354}{\textit{Low Temp. Phys.} \textbf{41}, 1024--1029 (2015).}

\bibitem{Bozhko2017} Bozhko, D.A. \textit{et al.}
Bottleneck accumulation of hybrid magnetoelastic bosons, 
\textit{Phys. Rev. Lett.} \textbf{118}, 237201 (2017).
\href{https://doi.org/10.1103/PhysRevLett.118.237201}{\textit{Phys. Rev. Lett.} \textbf{118}, 237201 (2017).}

\bibitem{Kreil2018} 
Kreil, A.J.E. \textit{et al.}
From kinetic instability to Bose-Einstein condensation and magnon supercurrents.
\textit{Phys. Rev. Lett.} (2018).

\bibitem{Dzyapko2017} Dzyapko, O. \textit{et al.} 
Magnon-magnon interactions in a room-temperature magnonic Bose-Einstein condensate. 
\href{https://doi.org/10.1103/PhysRevB.96.064438}{\textit{Phys. Rev. B} \textbf{96}, 064438 (2017).}

\bibitem{Pitaevskii2003} Pitaevskii, L.P. \& Stringari, S. 
\textit{Bose-Einstein condensation} (Oxford University Press, New York, 2003).


\bibitem{Nakata2014} Nakata, K., van Hoogdalem, K.A., Simon, P. \& Loss, D.
Josephson and persistent spin currents in Bose-Einstein condensates of magnons.
\href{https://doi.org/10.1103/PhysRevB.90.144419}{\textit{Phys. Rev. B} \textbf{90}, 144419 (2014).}
	
\bibitem{Chumak2015}  Chumak, A.V., Vasyuchka, V.I., Serga, A.A., \& Hillebrands B.
Magnon spintronics.
\href{https://doi.org/10.1038/nphys3347}{\textit{Nature Phys.} \textbf{11}, 453--461 (2015).}

\bibitem{Schneider2018} 
Schneider, M. \textit{et al.}  
Bose-Einstein condensation of quasi-particles by rapid cooling. 
\href{https://arxiv.org/abs/1612.07305v2}{\textit{arXiv}:1612.07305v2.}



	}
\end{thebibliography}

\begin{thebibliography}{99}
	\setcounter{NAT@ctr}{50}
	
	{\fontfamily{phv}
	\fontsize{8pt}{9.0pt}
	\selectfont
	
	\bibitem{Glass1988} Glass, H.L. 
	Ferrite films for microwave and millimeter-wave devices. 
	\href{http://dx.doi.org/10.1109/5.4391}{\textit{Proc. IEEE} \textbf{76}, 151--158 (1988).}
		
	\bibitem{Buettner2000} B\"uttner, O. \textit{et al.}
	Linear and nonlinear diffraction of dipolar spin waves in yttrium iron garnet films observed by space- and time-resolved Brillouin light scattering.
	\href{https://doi.org/10.1103/PhysRevB.61.11576}{\textit{Phys. Rev. B} \textbf{61}, 11576--11587 (2000).}
	
	\bibitem{Sandercock1981} Lindsay, S.M., Anderson, M.W. \& Sandercock, J.R. Construction and alignment of a high performance multipass vernier tandem Fabry-P\'{e}rot interferometer. 
	\href{https://doi.org/10.1063/1.1136479}{\textit{Rev. Sci. Instrum.} \textbf{52}, 1478--1486 (1981).}
	
	\bibitem{Mock1987} Mock, R., Hillebrands, B. \& Sandercock, J.R. Construction and performance of a Brillouin scattering set-up using a triple-pass tandem Fabry-P\'{e}rot interferometer. 
	\href{https://doi.org/10.1088/0022-3735/20/6/017}{\textit{J. Phys. E: Sci. Instrum.} \textbf{20}, 656--659 (1987).}
	
	\bibitem{Hillebrands1999} Hillebrands, B. Progress in multipass tandem Fabry-Perot interferometry: I. A fully automated, easy to use, self-aligning spectrometer with increased stability and flexibility.
	\href{https://doi.org/10.1063/1.1149637}{\textit{Rev. Sci. Instrum.} \textbf{70}, 1589--1598 (1999).}
	
	\bibitem{Sandweg2010} Sandweg, C.W. \textit{et al.}
	Wide-range wavevector selectivity of magnon gases in Brillouin light scattering spectroscopy,
	\href{https://doi.org/10.1063/1.3454918}{\textit{Rev. Sci. Instrum.} \textbf{81}, 073902 (2010).}
	
	\bibitem{COMSOL} https://www.comsol.com/
	
	\bibitem{Mihalceanu2018} Mihalceanu, L. \textit{et al.}
	Temperature dependent relaxation of dipolar-exchange magnons in yttrium-iron-garnet films. 
	\textit{Phys. Rev. B} \textbf{97}, 214405 (2018).
	
	}
\end{thebibliography}
\end{document}